\documentstyle[epsf,12pt,fleqn]{article}

\begin{document}

\begin{flushright}
 TIT/HEP-420/NP
\end{flushright}

\begin{center}
{\Large \bf
 Renormalization Group Approach to the $O(N)$ Linear Sigma Model
 at Finite Temperature} \par
\vskip 10mm
T. Umekawa\footnote{E-mail address: umekawa@th.phys.titech.ac.jp},
K. Naito$^A$ 
 and 
M. Oka\\ 
{\it Department of Physics, Tokyo Institute of Technology,} \\
{\it Meguro, Tokyo 152-8551, Japan} \\
{\it $^A$Radiation Laboratory, the Institute of Physical 
and Chemical Research (Riken),} \\
{\it Wako, Saitama 351-0198, Japan}
\end{center}
\vskip 5mm
\begin{abstract}
\baselineskip 1.5pc
 The Wilsonian renormalization group (RG) method is applied to 
 finite temperature systems for the study of 
 non-perturbative methods in the field theory. 
 We choose the $O(N)$ linear sigma model as the first step. 
 Under the local potential
 approximation, we solve the Wilsonian RG equation as 
 a non-linear partial differential equation numerically.
 The evolution of the domain is taken into account using the naive
 cut and extrapolation procedure. 
 Our procedure is shown to yield the correct solution obtained by
 the auxiliary field method
 in the large $N$ limit.
 To introduce thermal effects, we consider two schemes.
 One in which the sum of the Matsubara frequencies are taken before
 the scale is introduced is found to give more physical results.
 We observe a second order phase transition in both
 the schemes. The critical exponents are calculated and
 are shown to agree with the results from lattice calculations.
\end{abstract}
%
%
%
\section{Introduction} \label{SEC:H110412:1}
 Non-perturbative aspects of the 
 quantum field theory
 are very interesting and important.
 Study of low-energy hadron properties in QCD 
 is an example  
 that non-perturbative approaches are indispensable.
 We expect a phase transition at finite temperature which can be 
 understood only non-perturbatively.   
 Various non-perturbative approaches have been developed, such as the super-daisy summation
 of diagrams, the instanton approach, $1/N$ expansion, 
 $\epsilon$ expansion and so on. 
 The program, however, has not been accomplished yet.  
 Under this circumstance, the method of the 
 Wilsonian renormalization group (RG) equation\cite{WH73,WK74,Pol84} 
 receives much attention\cite{Mor93,EW94,ABBFTW95,AMSST96,AMSTT97,BJW97}.
 The idea is to include the quantum effects gradually from high energy 
 in order to obtain a low-energy (Wilsonian) effective action. 
 This RG equation is a non-linear evolution
 equation with respect to the scale $\Lambda$ 
 in the functional theory space.
 While $\Lambda$ is lowered, various 
 irrelevant operators are reproduced and finally we obtain the
 effective action in which all quantum effects are included.  
 This process can be achieved non-perturbatively.
 The concept of the Wilsonian RG equation is common to various 
 quantum field theories in the sense that it is model 
 independent in contrast to the 
 instanton or $1/N$ expansion approaches.
 A systematic approximation is also possible in the form of enlargement 
 of the functional theory space step by step. \par
 In spite of these advantages, 
 there are still technical difficulties in application.
 It is not easy to obtain the physical 
 quantities while it is very useful to see the phase 
 structure  and the critical phenomena. 
 This difficulty is especially serious in the symmetry broken phase. 
 In order to treat the bottom flatened effective potential, 
 which is naturally required in the quantum system, one must
 prepare a suitable functional theory space. 
 There is a dilemma that the expected potential is non-analytic while
 the RG equation is a partial differential equation which assumes 
 the smoothness of the solution. \par
 The aim of this paper is to challenge this dilemma. 
 We here propose to solve 
 the RG equation as a difference equation in the explicit scheme.
 In the symmetry broken phase, we set the domain in which the 
 solutions is continuous, and the domain size is modified at 
 each step of integration. 
 This approach may sound too naive but   
 we demonstrate the availability of such numerical
 treatment of the Wilsonian RG in physical applications. \par
 For this purpose we concentrate on the $O(N)$ linear sigma model. 
 This model is often  studied in other non-perturbative methods and 
 its physical application is also very interesting. 
 It can describe not
 only the spontaneous symmetry breaking but also 
 the symmetry restoration above a  critical temperature.
 Especially, the $O(N=4)$ symmetric model 
 is known to be equivalent to the chiral $SU_L(2)\times SU_R(2)$ effective 
 theory of QCD,
 and is widely expected to describe the chiral phase 
 transition in QCD. \par
 In Sect.2, we present our formulation of the Wilsonian RG equation and 
 the numerical method for solving the equation in the local potential 
 approximation. We demonstrate that our method works very well for the 
 large $N$ limit, for which the auxiliary field method yields the correct 
 answer. We apply our method for $N = 4$ and show that the symmetry 
 breaking is well represented in our method. \par
 In Sect.3, we apply our method to finite temperature $O(N)$ model. 
 We employ the imaginary time formalism. 
 When the sum of the Matsubara frequencies is considered, we  realize 
 two possible schemes of treating the scale dependence. 
 One is to cut-off the sum according to the scale change and the other
 is to take the sum before the scale is introduced. 
 These two schemes should give the same results provided that the 
 initial scale $\Lambda_0$ is  infinite. 
 However, in practice we choose a finite $\Lambda_0$ and the two schemes
 give different results. We examine both of them and in the end 
 show that the second scheme is more appropriate. \par
 The results for the $O(N)$ sigma model are presented.
 We obtain the second
 order phase transition to the chiral restored phase
 in the above two schemes. It is shown that the 
 four dimensional system near the critical temperature 
 in the first scheme reduces to a three dimensional system 
 at zero temperature. 
 This indicates the first scheme corresponds to taking the high 
 temperature limit even near the critical temperature.
 On the other hand, in the second scheme we obtain more 
 physical results. The critical exponents are computed 
 and are shown to agree very well to 
 the results in the lattice calculation. \par
 The conclusion is given in Sect.4. 
%
%
%
\section{Wilsonian RG Equation} \label{SEC:H110412:2}
 The lagrangian of the $O(N)$ linear sigma model we consider here 
 is given by 
\begin{equation}
 {\cal L} := \frac{1}{2} (\partial_\mu \phi)^2 + \frac{\mu^2}{2} \phi^2
  + \frac{\lambda}{8N} (\phi^2)^2
  \label{AEQ:H110412:1}
\end{equation}
 in the Euclidean notation $\mu=1,2,3$ and $4$. $\phi$ denotes the $N$
 component column vector $\phi:=(\phi_1,\phi_2,\cdots,\phi_N)^T$.\par
 The Wilsonian RG equation with a sharp cut-off scheme is called 
 Wegner--Houghton (WH) equation. The WH equation in the $O(N)$ 
 linear sigma model is derived in 
 Ref.\cite{AMSST96} and is given by
\begin{equation}
 -\Lambda \frac{d \Gamma_\Lambda}{d \Lambda} 
  = \frac{\Lambda}{2} \int_{|p|=\Lambda} d^Dp \left\{
 -{\rm Tr}{\rm Ln} \left( \frac{1}{\Lambda^2} 
 \frac{\delta^2 \Gamma_{\Lambda}}{\delta \phi_p \delta \phi_{-p}}
 \right) +
 \frac{\delta \Gamma_\Lambda}{\delta \phi_p}
 \left( \frac{\delta^2 \Gamma_\Lambda}{\delta \phi_p \delta \phi_{-p}}
 \right)^{-1} \frac{\delta \Gamma_\Lambda}{\delta \phi_{-p}} \right\}
  \label{AEQ:H110412:2}
\end{equation}
 in $D$ dimensional space-time.
 $\Gamma_{\Lambda}[\phi]$ denotes a scale dependent 
 effective action at the scale $\Lambda$ above which
 the quantum fluctuation is included.
 Thus at $\Lambda=0$ the scale dependent effective action agrees
 with the conventional effective action $\Gamma[\phi]$.
 The integral of r.h.s. is the shell ($|p|=\Lambda$) integral
\begin{equation}
 \int_{|p|=\Lambda} d^D p := \int \frac{d^D p}{(2\pi)^D} 
 \delta(|p|-\Lambda). 
 \label{AEQ:H110412:2a}
\end{equation} 
 The first and the second 
 terms in r.h.s. of Eq.$(\ref{AEQ:H110412:2})$ give ring and  dumbbell
 diagrams respectively.
 For a given initial condition $\Gamma_{\Lambda_0}[\phi]$ at the scale
 $\Lambda_0$, the effective action 
 $\Gamma_{\Lambda}[\phi]$ at any scale $\Lambda$ 
 is obtained by solving the WH equation.
 As the initial condition for Eq.$(\ref{AEQ:H110412:2})$ we
 assume that the effective action 
 $\Gamma_{\Lambda}[\phi]$ reduces at high $\Lambda_0$ 
 to the classical action $S[\phi]$. \par
 In practice, to solve the equation $(\ref{AEQ:H110412:2})$, suitable
 approximation
 is necessary. We employ the local potential approximation (LPA) 
 which assume the functional space as
\begin{equation}
 \Gamma_\Lambda[\phi] = \int d^D x \left\{ 
 \frac{1}{2}(\partial_\mu \phi)^2 + V_\Lambda(\phi) \right\}.
 \label{AEQ:H110414:1}
\end{equation}
 Under this approximation, the dumbbell diagram does not 
 contribute and the WH equation reads
\begin{equation}
 \frac{ d V_t(x) }{ d t } = \frac{A_D \Lambda^D}{2} 
 \left\{ (N-1)\ln\left(1 + \frac{1}{\Lambda^2} 
 \frac{1}{x} \frac{\partial V_t}{\partial x} \right)
 + \ln \left( 1 + \frac{1}{\Lambda^2} 
 \frac{\partial^2 V_t}{\partial x^2}\right) \right\}
 \label{AEQ:H110414:2}
\end{equation}
 where $\Lambda := \Lambda_0 \exp(-t)$, $x := \sqrt{ \phi^2 }$, 
\begin{equation}
 A_D := \frac{ \pi^{-D/2} 2^{1-D} }{ \Gamma(D/2) }
 \label{AEQ:H110513:1}
\end{equation}
and $V_t(x)$ is a scale dependent effective potential. \par
 We solve this equation numerically as a difference equation
 in the explicit scheme
\begin{eqnarray}
 \lefteqn{ \frac{V^{j+1}_i-V^j_i}{\Delta t} 
 = \frac{A_D \Lambda^D_j}{2} \Bigg\{ (N-1)
 \ln\left(1+\frac{1}{\Lambda^2_j} \frac{1}{x_i} 
 \frac{V^j_{i+1}-V^j_{i-1}}{2 \Delta x} \right) } \nonumber \\
 & & {}  + \ln\left( 1 + \frac{1}{\Lambda_j^2}
 \frac{ V^j_{i+1}-2 V^j_i + V^j_{i-1}}{ (\Delta x)^2 } \right) \Bigg\}
\label{AEQ:H110414:3}
\end{eqnarray}
 where $t_j := j \Delta t$, $\Lambda_j := \Lambda_0 \exp(-t_j)$, 
 $x_i := i \Delta x$ and $V^{j}_i:=V_{t_j}(x_i)$ 
 with $j=0,\cdots,j_{\rm max}$ and $i=0,\cdots,i_{\rm max}$. 
 This explicit scheme suggests us that a large 
 $j_{\rm max}$ is necessary for a large $i_{\rm max}$.
 Furthermore there is a subtle problem in solving 
 Eq.$(\ref{AEQ:H110414:3})$,
 that is , we have to specify the domain ${\cal D}$ of $x$.
 In the broken phase, the field variable can not take all the values 
 because the field variable $\phi$ is not an analytic
 function of the source $J$ around the origin $\phi = 0$. 
 Numerically the arguments of logarithm become  
 zero or negative for $x$ with $|x|\le \, ^\exists \! a$ at some $t_j$.
 In such cases, we obtain the effective potential
 at $t_{j+1}$ only by  the information of the 
 region $x$ with $|x|> a$.
 Thus, we consider the domain ${\cal D} = \{ x \,|\, |x|>a \}$.
 On the other hand, when the arguments of the logarithm grows up from 
 zero at the end point $x=a$ of the domain ${\cal D}$ at $t_{j}$,  
 we extrapolate the arguments as a smooth function of $x$ 
 using vicinal four points by
\begin{eqnarray}
 f(x_{m-1}) & \approx & f(x_m) - f'(x_m) \Delta x \nonumber \\
 & \approx & f(x_m) - \left( f'(x_{m+1}) - f''(x_{m+1}) \Delta x \right)
 \Delta x \nonumber \\
 & \approx & f(x_m) - \left( \frac{ f(x_{m+2}) - f(x_m) }{ 2 \Delta x } 
 - \frac{ f(x_{m+2}) - 2 f(x_{m+1}) + f(x_m) }{ \Delta x } \right)
 \Delta x \label{AEQ:H110419:1}
\end{eqnarray}
 where $x_m$ is the nearest neighbor to $a$
 and we enlarge the domain ${\cal D}$ at $t_{j+1}$. 
 The numerical result seems 
 to be almost independent of the extrapolation procedure if we take 
 sufficiently large  $i_{\rm max}$ (and $j_{\rm max}$). \par
 We choose the mass unit as $\Lambda_0$ so that $\Lambda_0=1$,
 concentrate on $D=4$, and set $\hat{\mu^2} = \mu^2/\Lambda_0^2 = -0.6$
 and $\lambda/8N = 4$ so as to bring out the spontaneous 
 symmetry breaking.   
 We employ the numerical parameters $i_{\rm max}=400$ 
 and $j_{\rm max}=10^5$.\par
\begin{figure}[tbp]
 \centerline{ \epsfxsize=11cm \epsfbox{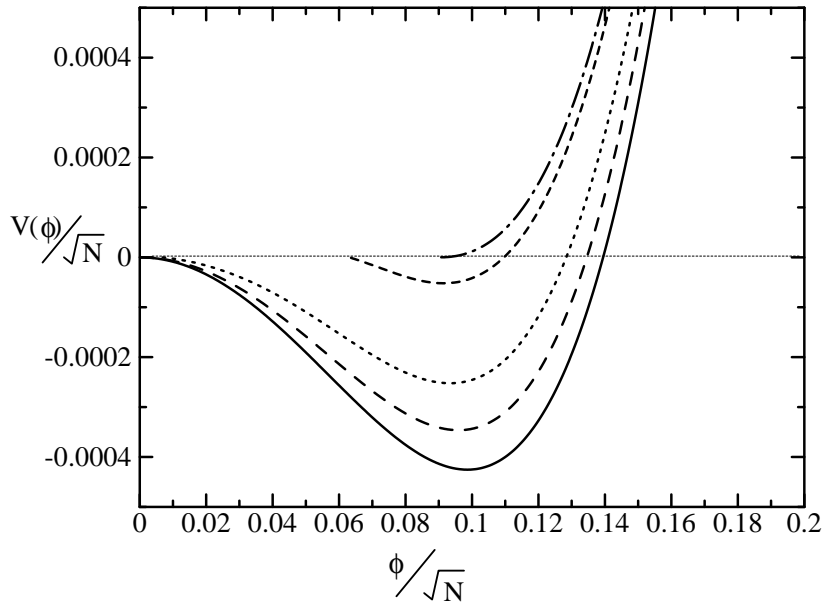} }
 \caption{
  The solutions of the  Wilsonian RG equation for the large $N$ limit.
  The curves correspond to $t=0.0,0.2,0.5,1.0,9.9$ from bottom to top 
  respectively.
 }
  \label{FIG:rg-lgbrk}
\end{figure}

 First we consider the large $N$ limit,
 where the LPA WH equation $(\ref{AEQ:H110414:2})$
 reduces to
\begin{equation}
 \frac{ d V_t(x) }{ d t } = \frac{A_D \Lambda^D}{2} 
  N \ln\left(1 + \frac{1}{\Lambda^2} 
 \frac{1}{x} \frac{\partial V_t}{\partial x} \right).
  \label{AEQ:H110414:4}
\end{equation}
 Eq.$(\ref{AEQ:H110414:4})$ is solved for the initial condition
\begin{equation}
 V_{t=0}(x) = \frac{\mu^2}{2}x^2 + \frac{\lambda}{8N} x^4.
 \label{AEQ:H110512:1}
\end{equation}
The result is shown in Fig.\ref{FIG:rg-lgbrk}. 
 For $t=0.0 \sim 0.5$, the effective potential becomes gradually 
 flat as the quantum correction is included. 
 This resembles the lattice result in Ref.\cite{MS96}. 
 For $t > 0.5$, however, the effective potential becomes 
 complex from the inside and the Domain develops. 
 As stated above, we do not use the values of such complex region
 in computing the next step. It is plausible that this region
 should be perfectly flattened as is indicated by the study 
 at $t \to \infty$ in Ref.\cite{MS96}.  
 It should be noted that the point of the potential minimum,
 which gives the vacuum expectation value of $\phi$, 
 and the end point of the domain become close to each other.
 Finally, at $t \to \infty$, these two points coincide
 and the effective potential defined conventionally is obtained.  
 \par
 It is well-known that 
 the exact effective potential can be obtained using the auxiliary
 field method in the large $N$ limit \cite{CJP74,KK75}.
  As a result, the scale dependent form of the 
 effective potential is given by
\begin{equation}
 V_t(x) = V_t(x,s) = \frac{1}{2} x^2 s - \frac{N}{2\lambda} s^2
 + \frac{N \mu^2}{\lambda} s + \frac{N}{2} \int_{\Lambda}^{\Lambda_0}
 \frac{d^4k}{(2\pi)^4} \ln (k^2 + s)
 \label{AEQ:H110414:5}
\end{equation} 
 where the auxiliary field $s$ is a function of $x$ and is given 
 by a solution of the saddle point condition
\begin{equation}
 \frac{ \partial V_t(x,s)}{\partial s } = 0.
 \label{AEQ:H110414:6}
\end{equation} 
 It is easy to see that Eq.$(\ref{AEQ:H110414:5})$ is the solution of 
 Eq.$(\ref{AEQ:H110414:4})$ with the initial condition 
 $(\ref{AEQ:H110512:1})$ \cite{AMSST99}. We find that our numerical 
 solutions in Fig.\ref{FIG:rg-lgbrk} agree almost completely to 
 the analytic solution Eq.$(\ref{AEQ:H110414:5})$. \par
 In plotting 
 Fig.\ref{FIG:rg-lgbrk}, we only take the region of $\phi$ where
 $V_t(x)$ is real. In fact, for each $t$ there is a critical value of 
 $x_c(t)=|\phi_c(t)|$ below which the solution of 
 Eq.$(\ref{AEQ:H110414:6})$ becomes complex.  
 In the present case, 
\begin{equation}
 x_c(t=\infty) = 2N \min_{s} \left\{ \frac{s-\mu^2}{\lambda} 
 - \frac{A_D}{4}
 \left( \Lambda_0^2-s\ln \frac{\Lambda_0^2+s}{s} \right)\right\}
 \label{AEQ:H110513:2}
\end{equation}
 This phenomenon occurs where the $O(N)$ symmetry is spontaneously broken
 and the system generates a non-zero vacuum expectation value of $\phi$.
 The magnitude of the vacuum condensate of $\phi$ is given by the critical
 value at $t\to \infty$ or $\Lambda\to 0$.\par
\begin{figure}[tbp]
  \centerline{ \epsfxsize=11cm \epsfbox{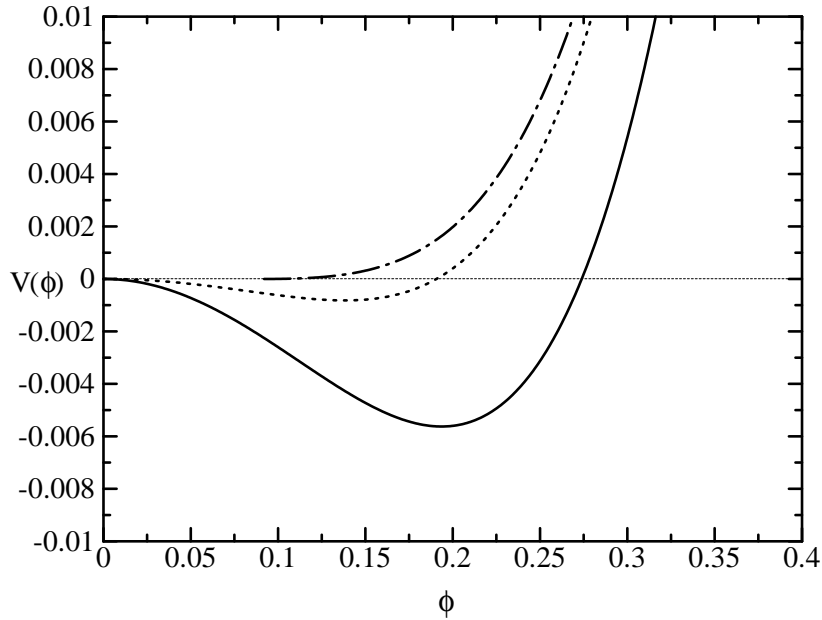} }
  \caption{
   Solution of the Wilsonian RG equation for $N=4$.
   The curves correspond to $t=0.0,0.5,10$ from bottom to top 
   respectively.
  }
  \label{FIG:rg-o4brk}
\end{figure}
\begin{figure}[tbp]
  \centerline{ \epsfxsize=8cm \epsfbox{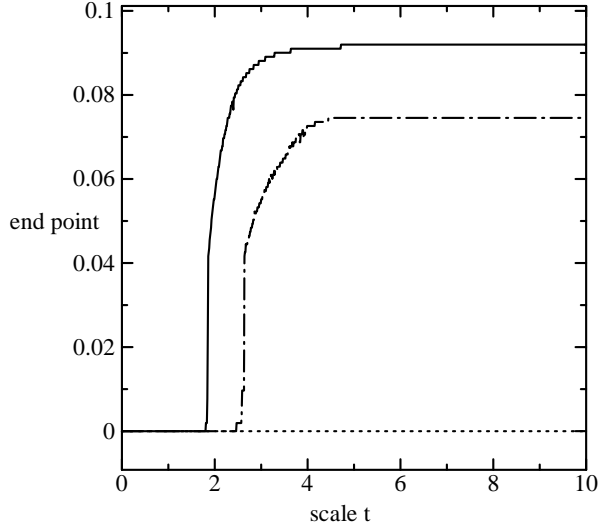} }
  \caption{
  The end points $a$ of the domain v.s. the scale variable $t$. 
  The curves correspond to the temperature $T=0.0,0.1,0.15$ 
  from top to bottom respectively. 
  \label{FIG:na-s2}
}
\end{figure}

 Next we consider the finite $N$ case.
 The O(4) result is shown in Fig.\ref{FIG:rg-o4brk}.
 Again the complex region appears. One sees that the qualitative behavior 
 of the effective potential is quite similar to that for the large $N$ limit 
 and that the calculation supports the spontaneous symmetry breaking. 
 The evolution of the domain by the scale is shown in Fig.\ref{FIG:na-s2}. 
 We see that the curves are monotonically increasing and saturate 
 at about $t \sim 5$ ($\Lambda \sim 0.007$). 
%
%
\section{Finite Temperature} \label{SEC:H110414:4}
 The introduction of the thermal effects is discussed in
 many text books \cite{KBDtextbook}.
 We use the imaginary time formalism in which the loop momentum
 integration is replaced by the sum of the Matsubara frequencies as
\begin{equation}
 \int \frac{d^D p}{(2\pi)^D} \Longrightarrow
 T \sum_{n} \int \frac{d^{D-1}p}{(2\pi)^{D-1}},
 \quad p_0 =  2n\pi T
 \label{AEQ:H110414:7}
\end{equation}
 where $T$ denotes the temperature.
 The system is now controlled by the new dimensional 
 parameter $T$. There are some candidates how to extend 
 the evolution equation in order to include the effect of $T$.
 We consider the following two schemes within the local 
 potential approximation. \par
\begin{center} {\bf Scheme I} \end{center}
 Firstly we consider the 4-dimensional spherical cut-off scheme. 
 For $T=0$, the formula 
\begin{eqnarray}
 \lefteqn{
 \Lambda \frac{1}{\delta \Lambda} \int_{\Lambda-\delta \Lambda\le p
 \le \Lambda} \frac{d^D p}{(2\pi)^D} = \Lambda \frac{1}{\delta
 \Lambda} \int_{\Lambda-\delta \Lambda}^{\Lambda} d p \, p^{D-1} 
 \int \frac{d\Omega_p}{(2\pi)^D} } \nonumber \\
 & \stackrel{\delta \Lambda\to 0}{\to} & \Lambda^D \int 
 \frac{d\Omega_p}{(2\pi)^D} = \int_{|p|=\Lambda} d p. 
 \label{AEQ:H110414:8}
\end{eqnarray}
 is used in the derivation of the WH equation $(\ref{AEQ:H110412:2})$. 
 If a function $f(p^2)$ depends only on $p^2$, then
 Eq.$(\ref{AEQ:H110414:8})$ reduces to
\begin{equation}
 \Lambda \frac{1}{\delta \Lambda} \int_{\Lambda-\delta 
 \Lambda\le p \le \Lambda} \frac{d^D p}{(2\pi)^D} f(p^2) 
 \stackrel{\delta \Lambda\to 0}{\to} \Lambda^D A_D f(\Lambda^2)
 \label{AEQ:H110414:9}
\end{equation}
 which is used as the prefactor in 
 r.h.s. of Eq.$(\ref{AEQ:H110414:2})$.
 Even at finite temperature, the integrand is still $f(p^2)$
 within LPA and therefore the shell integration can be performed as
\begin{eqnarray}
 \lefteqn{ 
 \Lambda \frac{1}{\delta \Lambda} \frac{1}{\beta}
 \sum \int_{\Lambda-\delta\Lambda\le |p| \le \Lambda}
 \frac{d^{D-1}p}{(2\pi)^{D-1}} f(p^2) } \nonumber \\
 & = & \frac{\Lambda}{\delta \Lambda} A_{D-1} \frac{1}{\beta}
 \sum \int_{\Lambda-\delta\Lambda\le\sqrt{(2\pi n/\beta)^2+p_r^2}
 \le \Lambda} dp_r p_r^{D-2} f\left( \left(\frac{2\pi n}{\beta}\right)^2
 + p_r^2 \right) \nonumber \\
 & = & \frac{\Lambda}{\delta \Lambda} A_{D-1}  
 \frac{1}{\beta} \sum_{|n|\le \Lambda\beta/(2\pi)}
 \int_{\sqrt{(\Lambda-\delta\Lambda)^2-(2\pi n/\beta)^2}}^{
 \sqrt{\Lambda^2 - (2\pi n/\beta)^2}} dp_r p_r^{D-2}
 f\left(\left(\frac{2\pi n}{\beta}\right)^2+p_r^2\right) \nonumber \\
 & \stackrel{\delta\Lambda\to 0}{\to } &
 \Lambda^{D-1} A_{D-1} \frac{1}{\beta}\sum_{|n|\le \Lambda\beta/(2\pi)}
 \left( 1-\left(\frac{2\pi n}{\Lambda \beta}\right)^2\right)^{
 \frac{D-3}{2}} f(\Lambda^2) \label{AEQ:H110414:10}
\end{eqnarray}
 where $\beta:=1/T$. Comparing Eq.$(\ref{AEQ:H110414:9})$ and
 Eq.$(\ref{AEQ:H110414:10})$, we conclude that the thermal effect
 is taken into account by replacing the prefactor 
 of r.h.s. of Eq.$(\ref{AEQ:H110414:2})$,
\begin{equation}
 A_D \Lambda^D \Longrightarrow A_{D-1} \frac{\Lambda^{D-1}}{\beta}
 \sum_{|n|\le \Lambda\beta / (2\pi) }
 \left( 1 - \left( \frac{2\pi n}{\Lambda \beta} \right) \right)
 ^{\frac{D-3}{2}}.
 \label{AEQ:H110414:11}
\end{equation}
Finally, we obtain the WH equation in the scheme I,
\begin{eqnarray}
 \lefteqn{ 
 \frac{ d V_t(x) }{ d t } =  
 \frac{A_{D-1} \Lambda^{D-1}}{2 \beta}
 \sum_{|n|\le \Lambda\beta / (2\pi) }
 \left( 1 - \left( \frac{2\pi n}{\Lambda \beta} \right) \right)
 ^{\frac{D-3}{2}} } \nonumber \\
 & & {} \times \left\{ (N-1)\ln\left(1 + \frac{1}{\Lambda^2} 
 \frac{1}{x} \frac{\partial V_t}{\partial x} \right)
 + \ln \left( 1 + \frac{1}{\Lambda^2} 
 \frac{\partial^2 V_t}{\partial x^2}\right) \right\}.
 \label{AEQ:H110414:15}
\end{eqnarray}
 Here it should be noted that the thermal loop contribution 
is summed up only for $|n|\le \Lambda_0\beta/2\pi$ in the scheme I. For 
$\Lambda_0\to \infty$, this scheme provides the full thermal loop 
correction, while for relatively small $\Lambda_0$, the summation 
is limited to low $|n|$ especially at high temperature.
Indeed, we will see later that the limited summation gives a 
fluctuation in the effective potential. In order to avoid this 
difficulty, we next consider the second scheme. 
\begin{center} {\bf Scheme II} \end{center}
  We consider the scheme in which the cut-off is introduced after
 the separation of the quantum and thermal effects. In this scheme, therefore,
 we perform the summation of all Matsubara frequencies. \par
 For simplicity, we focus only on LPA.
 Suppose that the effective potential 
 $V_t(\phi)$ is given at the scale $t$. 
 Then the effective potential $V_{t+\Delta t}(\phi)$ 
 is given by 
\begin{eqnarray}
 V_{t+\Delta t}(\phi) & = & V_t(\phi) + \frac{1}{2\beta}
 \sum \int \frac{d^{D-1} k}{(2\pi)^{D-1} } \ln \det \left[ k^2 + 
 \frac{\delta^2 V_t(\phi)}{\delta \phi^T \phi} \right] 
 \nonumber \\
& & {} + \mbox{ \vspace*{14pt} (2-loop and higher loop contributions) }
 \label{AEQ:H110414:12}
\end{eqnarray}
 where the integral is performed between $t$ and $t+\Delta t$.
 To specify the cut-off scheme we use the formula
\begin{eqnarray}
 \lefteqn{ 
 \frac{1}{\beta} \sum_{n} \int \frac{d^{D-1} k}{(2\pi)^{D-1} } 
 \ln( (2n\pi T)^2 + \vec{k}^2 + M^2 ) = \int \frac{d^D k}{(2\pi)^D } 
 \ln (k^2+M^2) } \nonumber \\
& & {} + \frac{2}{\beta} \int \frac{d^{D-1} k}{(2\pi)^{D-1} } 
 \ln( 1 - \exp(-\beta\sqrt{k^2+M^2}) ) \nonumber \\
 & = & \int_{\Lambda=0}^{\Lambda=\infty}d\Lambda
 \left\{ \frac{A_D \Lambda^{D-1} }{2} \ln(\Lambda^2+M^2)
 + \frac{A_{D-1} \Lambda^{D-2} }{\beta} 
 \ln(1-\exp(-\beta\sqrt{\Lambda^2+M^2}))
 \right\} \label{AEQ:H110414:13}
\end{eqnarray}
 where $\beta=1/T$ and $M^2$ is an arbitrary constant. In  r.h.s. 
 the quantum and thermal 
 effects are separated. Thus we introduce a sharp cut-off 
 of $\Lambda$ in Eq.$(\ref{AEQ:H110414:13})$.     
 Similar procedure can be performed for the 2-loop and higher
 loop contributions.
 The cut-off is introduced so that in the quantum correction part 
 the radial mode in the  4-dimensional integration
 is cut-off, while in the thermal correction part radial mode in the 
  3-dimensional integration
 which is left after the summation or the integration with respect to
 $k_0$ is cut-off.
 Taking the limit $\Delta t\to 0$ in Eq.$(\ref{AEQ:H110414:12})$,
 the 2-loop and higher loop contributions vanish.
 Therefore we conclude that the thermal effect is included by adding
\begin{eqnarray} 
 \lefteqn{ 
 \frac{A_{D-1} \Lambda^{D-1}}{\beta} \Bigg\{
 (N-1) \ln\left(1-\exp\left(-\beta\sqrt{\Lambda^2+ \frac{1}{x}
 \frac{\partial V_t}{\partial x}} \right) \right) } \nonumber \\
 & & {} +
 \ln\left(1-\exp\left(-\beta\sqrt{\Lambda^2+
 \frac{\partial^2 V_t}{\partial x^2}}  \right) \right) \Bigg\}
 \label{AEQ:H110414:14}
\end{eqnarray}
 to r.h.s. of Eq.$(\ref{AEQ:H110414:2})$.\par
 Finally we obtain the WH equation in the scheme II.
\begin{eqnarray}
 \lefteqn{ 
 \frac{ d V_t(x) }{ d t } = \frac{A_D \Lambda^D}{2} 
 \left\{ (N-1)\ln\left(1 + \frac{1}{\Lambda^2} 
 \frac{1}{x} \frac{\partial V_t}{\partial x} \right)
 + \ln \left( 1 + \frac{1}{\Lambda^2} 
 \frac{\partial^2 V_t}{\partial x^2}\right) \right\}
 } \nonumber \\
 & & {} + \frac{A_{D-1} \Lambda^{D-1}}{\beta} \Bigg\{
 (N-1) \ln\left(1-\exp\left(-\beta\sqrt{\Lambda^2+ \frac{1}{x}
 \frac{\partial V_t}{\partial x}} \right) \right) \nonumber \\
 & & {} +
 \ln\left(1-\exp\left(-\beta\sqrt{\Lambda^2+
 \frac{\partial^2 V_t}{\partial x^2}}  \right) \right) \Bigg\}.
 \label{AEQ:H110414:16}
\end{eqnarray}
 It is noted that neglecting the quantum effects (the first term in 
 r.h.s.), this evolution equation  coincides with the thermal 
 RG equation in Ref.\cite{BR99},
 which is derived in the real time formalism \cite{Ber98}, 
 up to an additional step function.\par
 In these two schemes,  
both Eqs.$(\ref{AEQ:H110414:15})$ and $(\ref{AEQ:H110414:16})$ reduce to
the zero temperature formula Eq.$(\ref{AEQ:H110414:2})$ in the limit $T\to 0$ 
and  they reduce in the limit $T\to \infty$ to the effective
$D-1$ dimensional form
\begin{equation}
 \frac{ d V_t(x) }{ d t } = \frac{A_{D-1} \Lambda^{D-1}}{2\beta} 
 \left\{ (N-1)\ln\left(1 + \frac{1}{\Lambda^2} 
 \frac{1}{x} \frac{\partial V_t}{\partial x} \right)
 + \ln \left( 1 + \frac{1}{\Lambda^2} 
 \frac{\partial^2 V_t}{\partial x^2}\right) \right\}.
 \label{AEQ:H110415:1}
\end{equation}
 The difference between the two schemes 
 come from the cut-off procedure.
 Taking $\Lambda_0\to \infty$, their effective potentials at
 $\Lambda=0$ coincide.
 In practice, however, we neglect the thermal effect above 
 $\Lambda_0$ in the first scheme and assume that $\Gamma_{\Lambda_0}$ 
 equals to the zero temperature action. Thus the two schemes give
  different results for finite $\Lambda_0$. \par
\begin{figure}[tbp]
  \centerline{ \epsfxsize=11cm \epsfbox{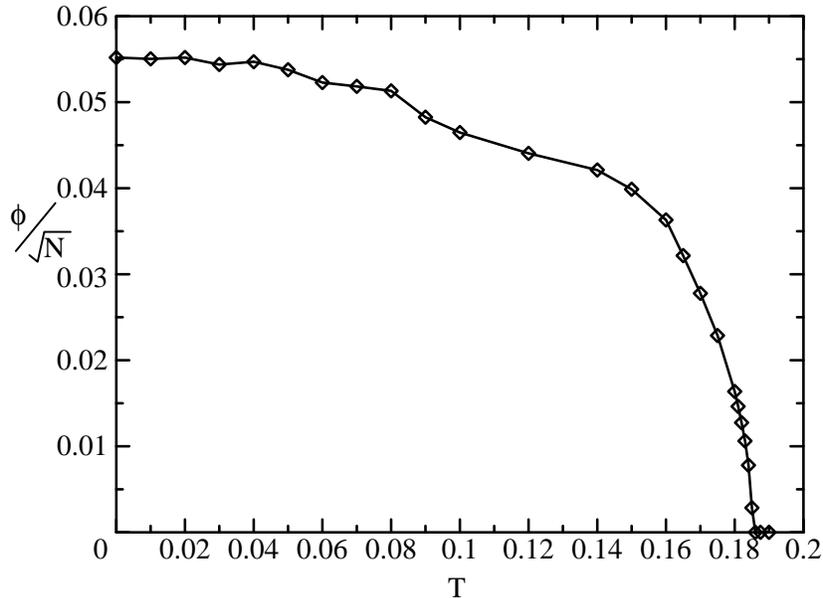} }
  \caption{Temperature  dependence of the condensation 
   $\phi$ as a function 
  of temperature $T$ in the scheme I for 
  the large $N$ limit. }
  \label{FIG:rg-lgcs1}
\end{figure}
\begin{figure}[tbp]
  \centerline{ \epsfxsize=11cm \epsfbox{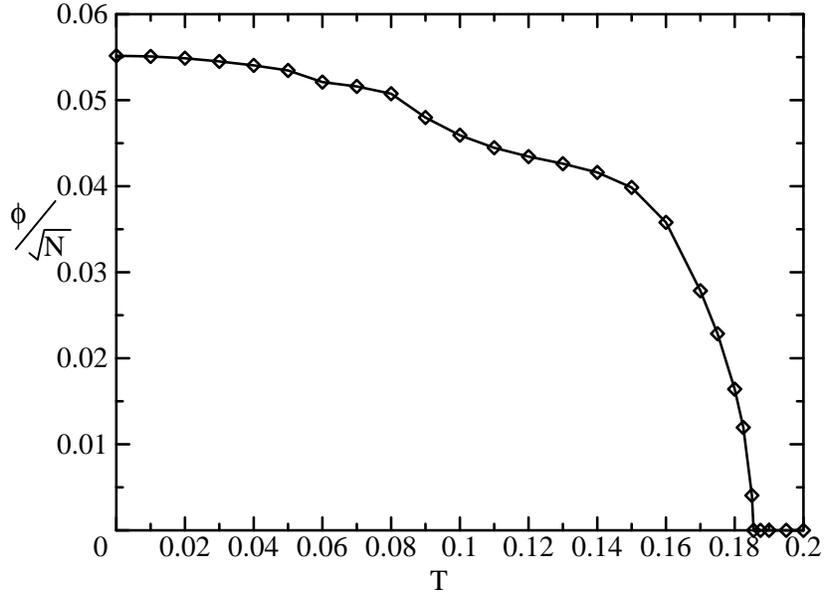} }
  \caption{The same as Fig.4 obtained 
  in the auxiliary field method. }
  \label{FIG:ctlgn}
\end{figure}
\begin{figure}[tbp]
  \centerline{ \epsfxsize=11cm \epsfbox{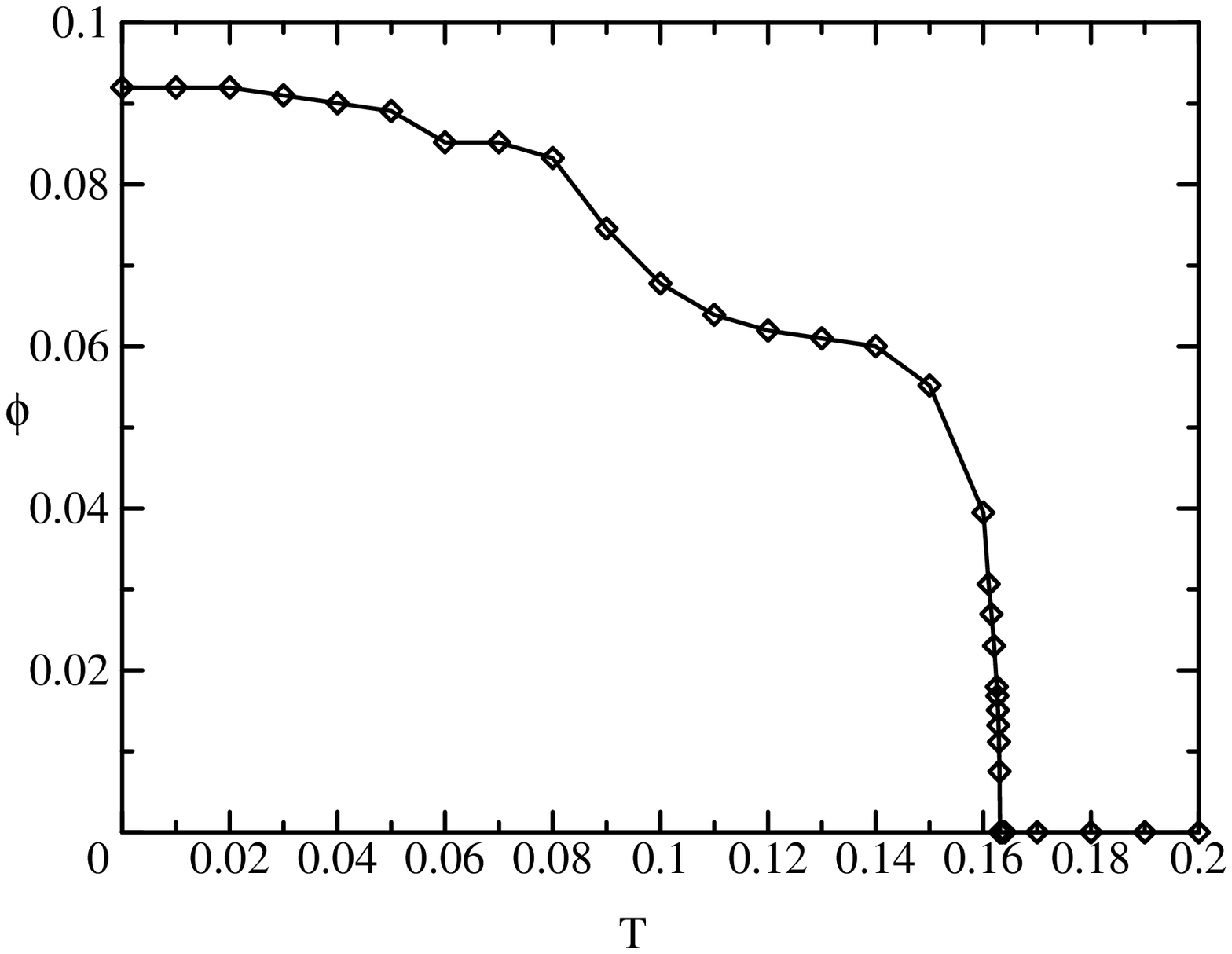} }
  \caption{The temperature  dependence of the condensation 
  $\phi$ in the scheme I for $N = 4 $. }
  \label{FIG:rg-o4cs1}
\end{figure}

 In the scheme I, the numerical results are as follows.
 Fig.\ref{FIG:rg-lgcs1} shows the vacuum expectation value of 
 $\phi$ as a function of the temperature $T$ in the large $N$ limit of 
 Eq.$(\ref{AEQ:H110414:15})$. Fig.\ref{FIG:ctlgn} shows the same 
 quantity calculated by using the auxiliary field method.
 As these two curves agree well, we conclude that our numerical
 approach is appropriate. The result  for $N=4$ is given in 
 Fig.\ref{FIG:rg-o4cs1}.  
 In these graphs, one sees unnatural steps. 
 They come from the summation in 
 Eq.$(\ref{AEQ:H110414:15})$.
 The summation is limited to a few terms when $T$ becomes comparable to
 $\Lambda_0/2\pi$ ($\approx 150$ [MeV]  for $\Lambda_0 = 1$ [GeV]). 
  As the vacuum expectation value is sensitive to the number 
 of the summation in the initial condition, there appear unnatural 
 steps seen in the results.
 Since the critical temperature $T_c=0.185$ shown in 
 Fig.\ref{FIG:rg-lgcs1}
 is larger than $\Lambda_0/2\pi$, the phase transition in this scheme I 
 is controlled by the high temperature formula Eq.$(\ref{AEQ:H110415:1})$. 
 At low temperature, we do not see such steps and the results agree 
 very well with those obtained in our second scheme. 

\begin{figure}[tbp]
  \centerline{ \epsfxsize=8cm \epsfbox{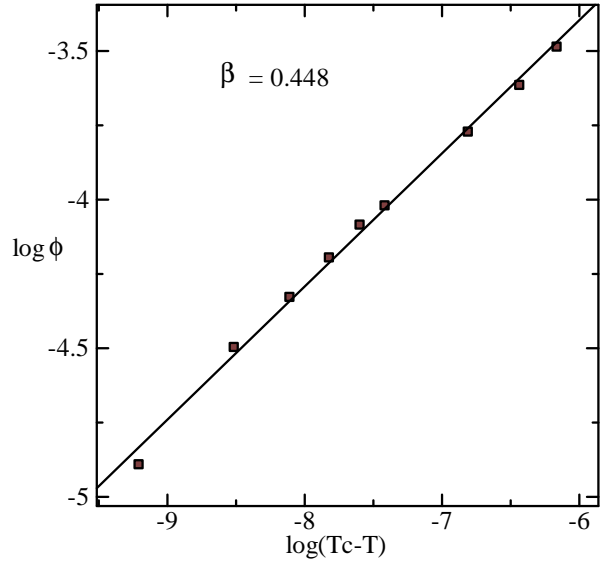} }
  \caption{
  The critical exponent $\beta$ for the result of the Wilsonian 
  RG method for $N=4$ in the scheme I.
}
  \label{FIG:s1b}
\end{figure}

\begin{figure}[tbp]
  \centerline{ \epsfxsize=8cm \epsfbox{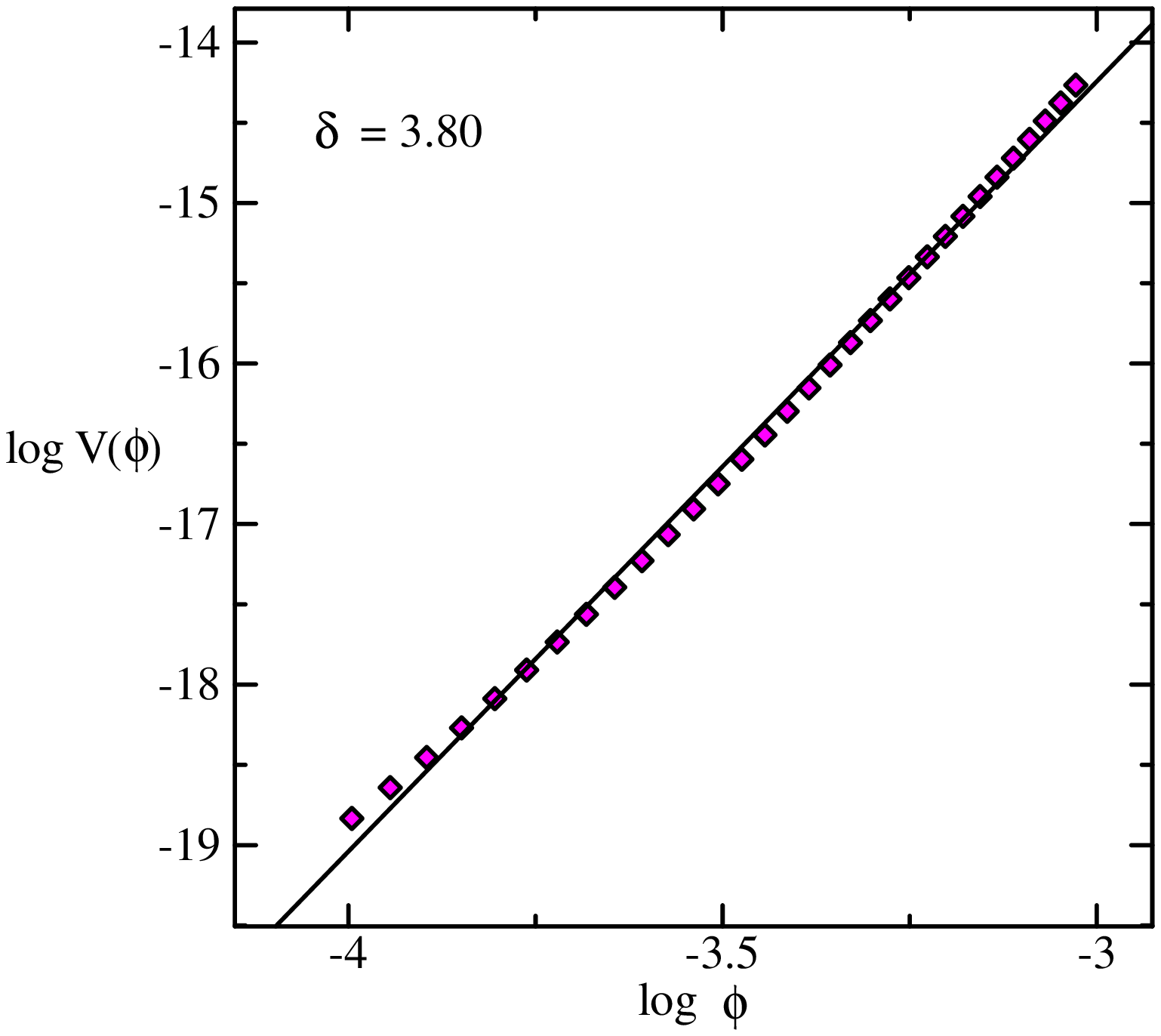} }
  \caption{
  The critical exponent $\delta$  for $N=4$
  in the scheme I. 
}
  \label{FIG:s1d}
\end{figure}

\begin{figure}[tbp]
  \centerline{ \epsfxsize=8cm \epsfbox{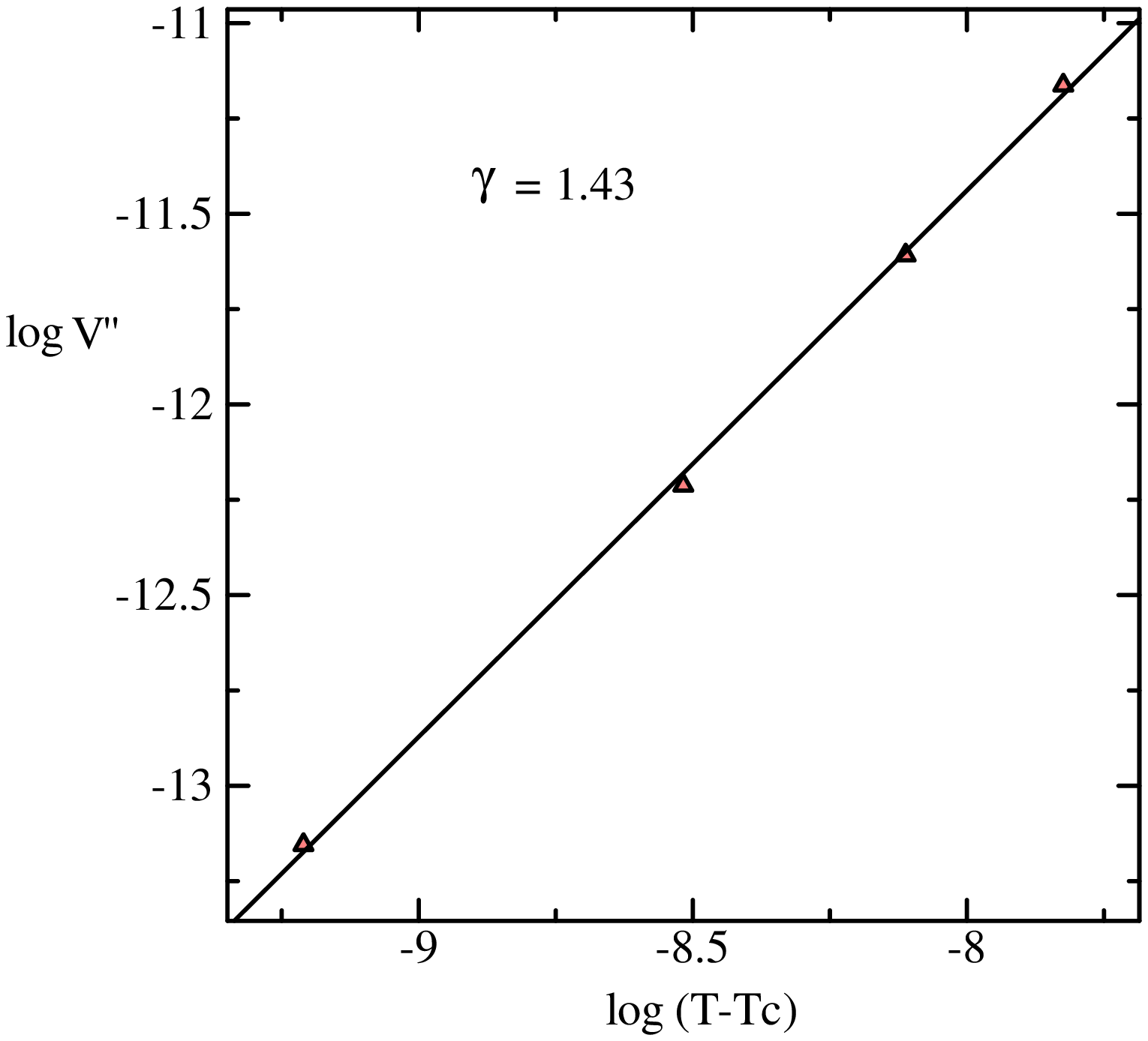} }
  \caption{
  The critical exponent $\gamma$  for $N=4$ 
  in the scheme I. 
  }
  \label{FIG:s1g}
\end{figure}

 In Fig.\ref{FIG:s1b}, we plot 
 $\log(\phi)$ as a function of $\log(T_c-T)$, 
 where $T_c$ denotes the critical temperature.
 After fitting the curve to the linear function using the least 
 square method,
 we obtain the critical exponent $\beta=0.448$, defined by
\begin{equation}
 \phi \propto |T_c-T|^\beta.
 \label{AEQ:H110510:1}
\end{equation}
 We also obtain the critical exponent $\delta=3.80$ from 
Fig.\ref{FIG:s1d}, where $\delta$ is defined by
\begin{equation}
 V \propto |\phi|^{\delta+1}
 \label{AEQ:H110517:1}
\end{equation}
at $T=T_c$.
In the same way we obtain the critical exponent $\gamma=1.43$ from 
Fig.\ref{FIG:s1g}, where $\gamma$ is defined by
\begin{equation}
 \frac{d^2 V}{d \phi^2} \propto |T_c-T|^{\gamma}.
 \label{AEQ:H110517:2}
\end{equation}
\par
 These results are summarized in Table \ref{TBL:H110517:1}.
 From Eqs.$(\ref{AEQ:H110510:1})$-$(\ref{AEQ:H110517:2})$, we
 obtain a scaling relation among the exponents,
\begin{equation}
 \gamma = \beta ( \delta - 1 ). 
 \label{AEQ:H110526:4}
\end{equation}
 Our numerical results yield $\beta(\delta-1)/\gamma=0.877$, 
 which is not far from $1$. 
 \par
\begin{figure}[tbp]
  \centerline{ \epsfxsize=11cm \epsfbox{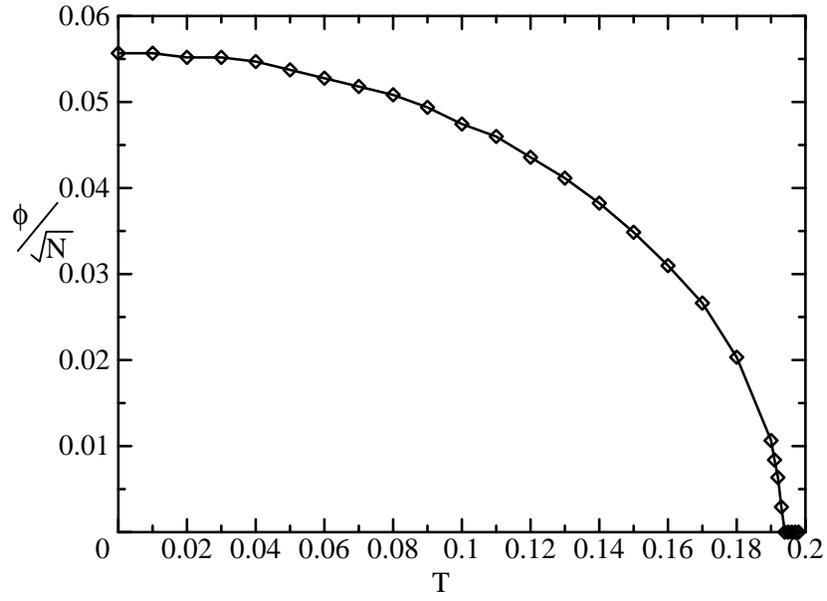} }
  \caption{The temperature  dependence of the condensation 
  $\phi$ in the scheme II for the large $N$ limit. }
  \label{FIG:rg-lgcs2}
\end{figure}
\begin{figure}[tbp]
  \centerline{ \epsfxsize=11cm \epsfbox{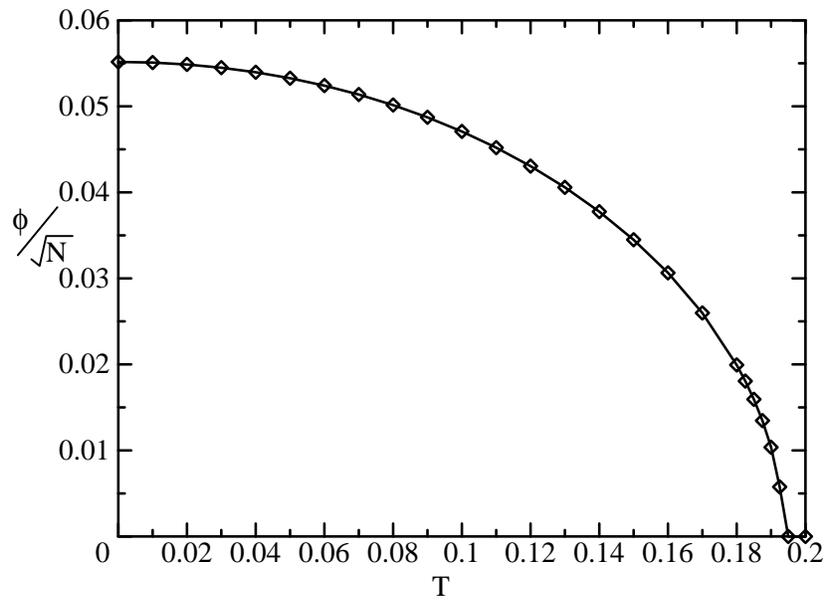} }
  \caption{The same as Fig.10 in the auxiliary field method.  }
  \label{FIG:ctlgn2}
\end{figure}
\begin{figure}[tbp]
  \centerline{ \epsfxsize=11cm \epsfbox{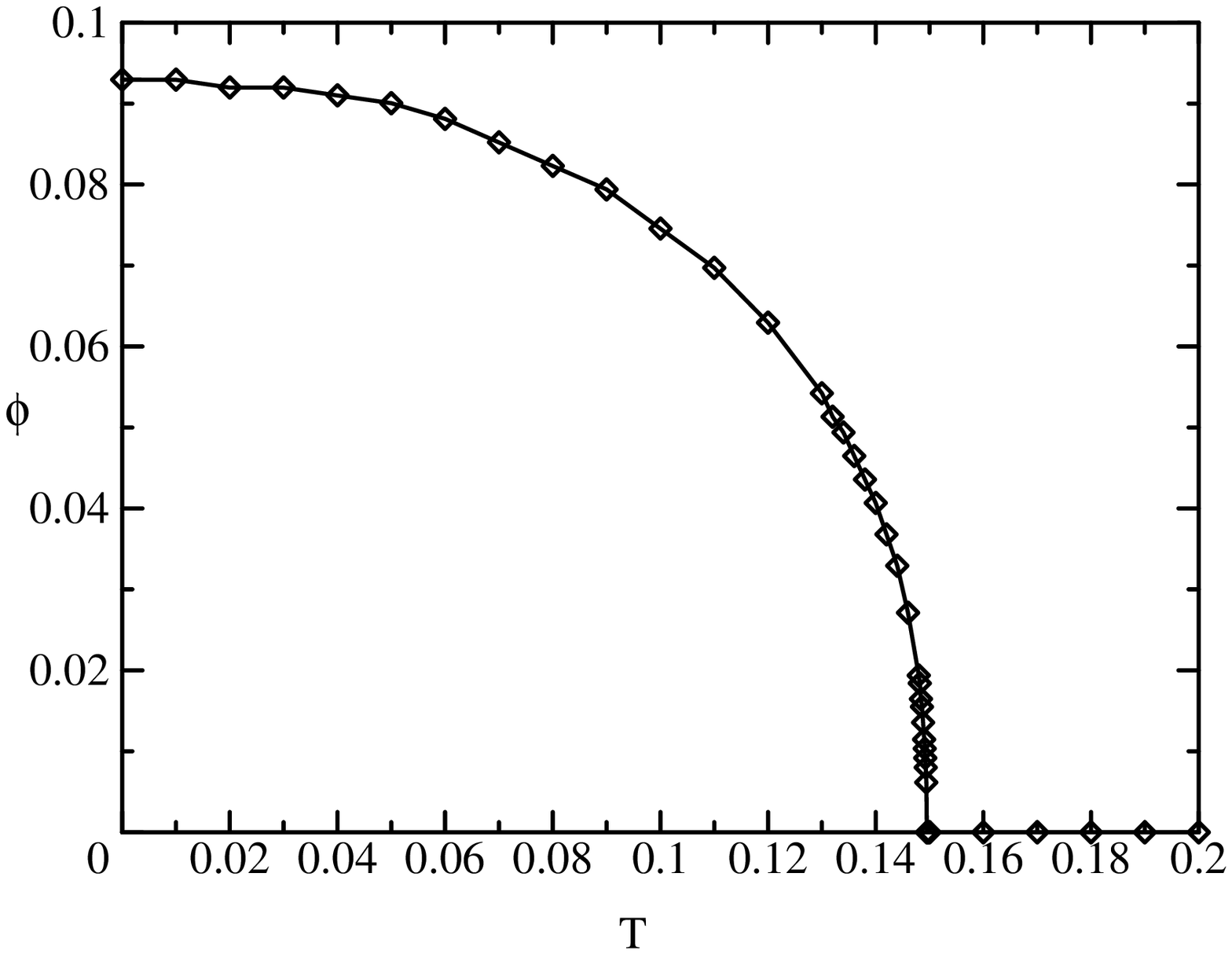} }
  \caption{The temperature dependence of the condensation 
  $\phi$ in the scheme II 
  for $N = 4$. }
  \label{FIG:rg-o4cs2}
\end{figure}
\begin{figure}[tbp]
  \centerline{ \epsfxsize=8cm \epsfbox{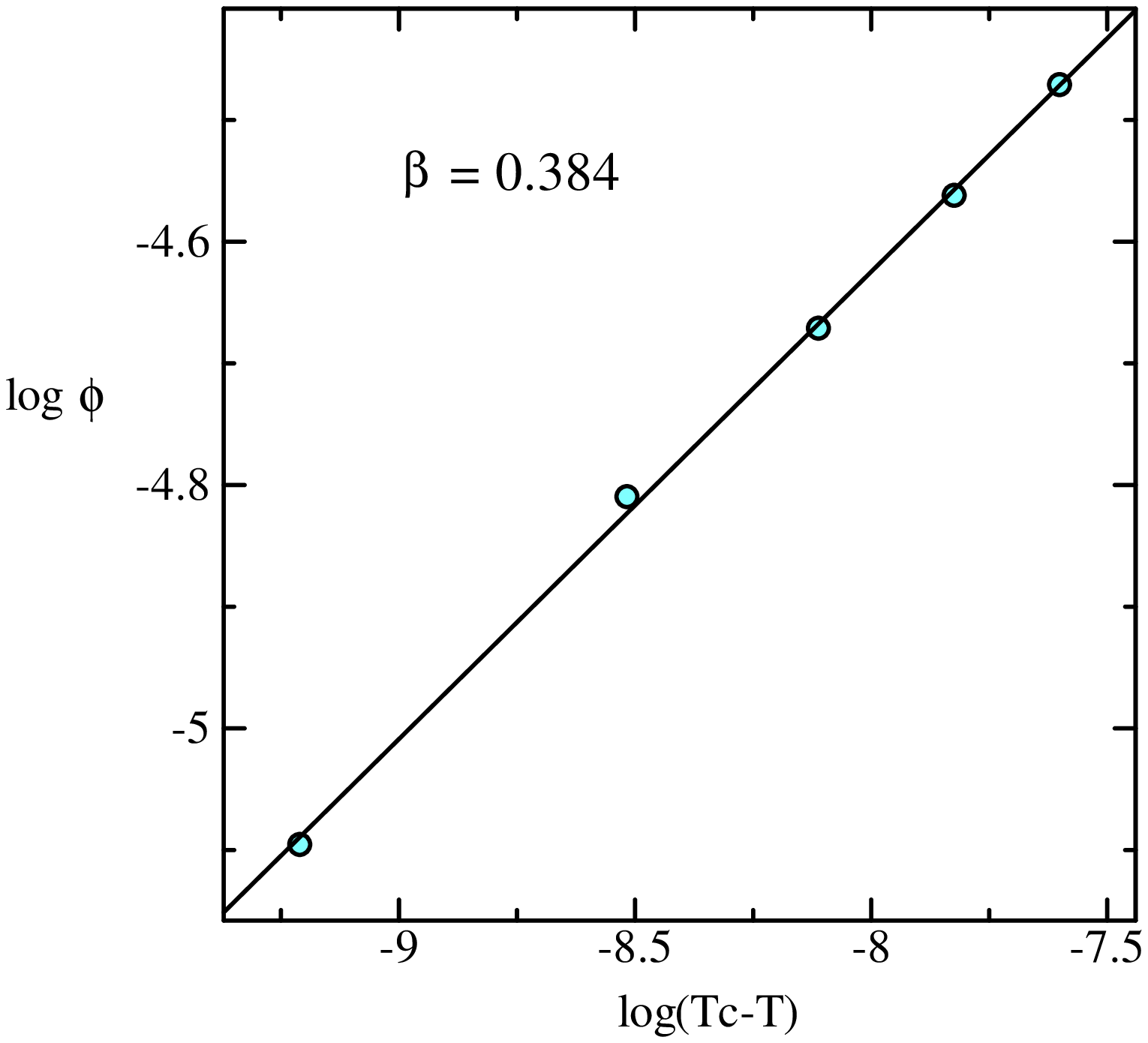} }
  \caption{
  The critical exponent $\beta$ for $N=4$ in the scheme II.
}
 \label{FIG:s2b}
\end{figure}
\begin{figure}[tbp]
  \centerline{ \epsfxsize=8cm \epsfbox{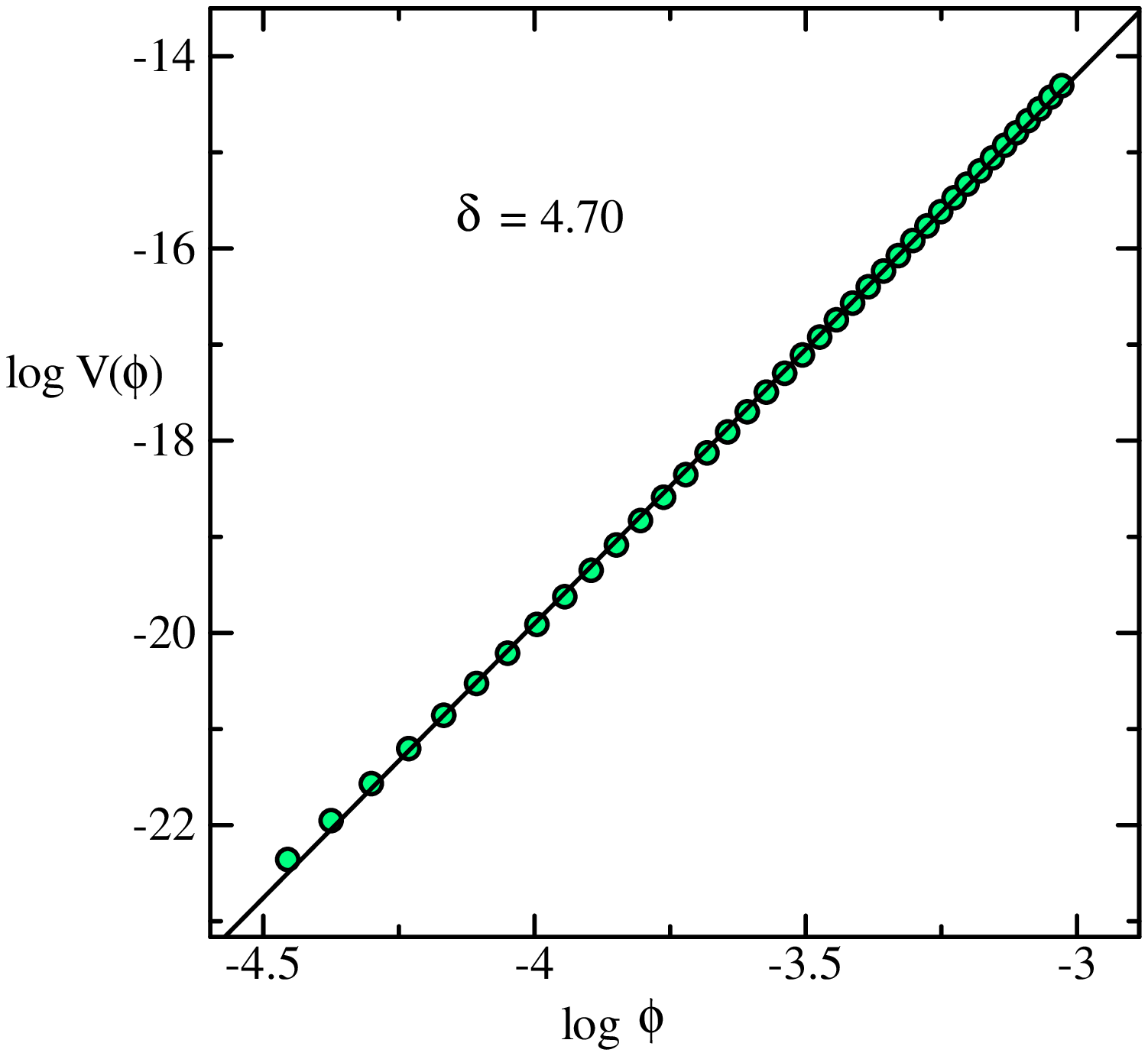} }
  \caption{
  The critical exponent $\delta$ for $N=4$ in the scheme II.
  \label{FIG:s2d}
}
\end{figure}
\begin{figure}[tbp]
  \centerline{ \epsfxsize=8cm \epsfbox{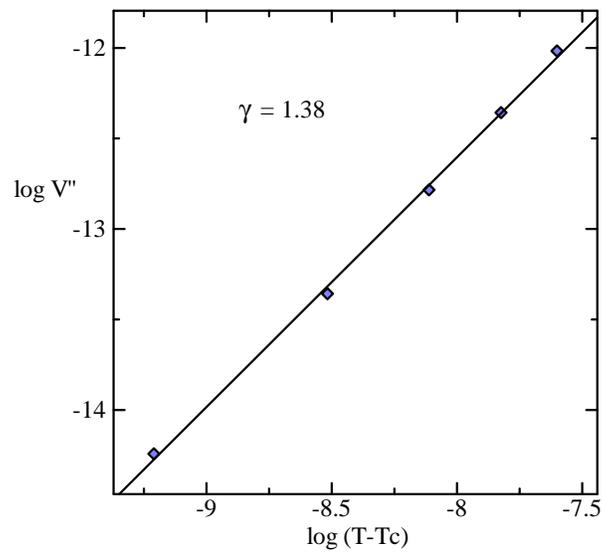} }
  \caption{
  The critical exponent $\gamma$ for $N=4$ in the scheme II.
} 
  \label{FIG:s2g}
\end{figure}

 Numerical results of the scheme II are shown in 
 Figs.\ref{FIG:rg-lgcs2}-\ref{FIG:rg-o4cs2} and
 Table \ref{TBL:H110517:1}.
 In this scheme the summation of the infinite number
 of the Matsubara frequencies is performed.  
 Therefore in the scheme II the unnatural steps seen in the 
 scheme I do not appear. These curves again show 
 the second order chiral phase transition. 
 We estimate the critical exponent $\beta=0.384$ 
 from Fig.\ref{FIG:s2b}.  This value agrees with the 
 value $0.384$ obtained by the Monte-Calro calculation in 
 Ref.\cite{KK95}, 
 where the $O(N)$ Heisenberg model in three dimension is used.
 We also obtain the critical exponent $\delta=4.70$
 from Fig.\ref{FIG:s2d}. The value given in Ref.\cite{KK95}
 is $\delta=4.85$.
 In the same way, we obtain the critical exponent $\gamma=1.38$ 
 from Fig.\ref{FIG:s2g}. The value in Ref.\cite{KK95} 
 is $\gamma=1.48$ which is also 
 close to ours.
 These critical exponents satisfy the scaling relation very well,
 giving $\beta(\delta-1)/\gamma=1.03$. 
\begin{table}[htbp]
\begin{center}
\begin{tabular}{c|cccc}  \hline \hline
           &  $\beta$ & $\delta$ & $\gamma$ & $\beta(\delta-1)/ \gamma$ \\ \hline
 Scheme I  &  $0.448$ & $3.80$ & $1.43$ & $0.877$ \\ \hline
 Scheme II & $0.384$ & $4.70$ & $1.38$ & $1.03$ \\ \hline 
 Monte-Calro\cite{KK95} & $0.384$ & $4.85$ & $1.48$ & --- \\ \hline \hline
\end{tabular} 
\end{center}  
 \caption{Critical Exponents}
 \label{TBL:H110517:1}
\end{table}    
%
%
\section{Conclusion} \label{SEC:H110414:3}
%
%
 We have applied the Wilsonian RG method as the 
 non-perturbative calculation method in the field theory to 
 the $O(N)$ linear sigma model and calculated the low energy
 effective potential in the local potential approximation. 
 We have introduced a numerical recipe to define the domain 
 for the effective potential and have tested the method in 
 the large $N$ limit. 
 The results coincide with the solutions obtained by the 
 auxiliary field method very well. 
 This means that our numerical treatment of the Wilsonian 
 RG equation is sufficiently good in this situation. 
 We have also shown that the $N=4$ effective potential behaves 
 similarly as those in the large $N$ limit. 
\par
 We have applied our approach to the finite temperature 
 system to see the chiral restoration phase transition. 
 We have employed the imaginary time formulation using the sum of 
 the Matsubara frequencies. 
 When we take the sum of the Matsubara frequencies, we consider 
 two schemes. 
 In the scheme I, we use the 4-dimensional spherical cut-off. 
 Then the sum of the Matsubara frequencies is also limited accordingly. 
 In the scheme II, we separate the quantum and thermal 
 effects, before the momentum integral is cut-off. 
 These two schemes should agree with each other if 
 the initial $\Lambda_0$ is infinite. 
 In the scheme I, however, we assume that the thermal effect above 
 $\Lambda_0$ is negligible and $\Gamma_{\Lambda_0}$ equals to the 
 zero temperature action.
 Thus the two schemes give different results for
 finite $\Lambda_0$. 
 Both of them show the second order phase 
 transition.
 But the results of the scheme I have unnatural steps.
 These steps arise in the course of transfer from the 
 four dimensional system to the three dimensional one at
 high temperature. On the other hand 
 we have found that the scheme II gives more natural and 
 physical results. 
 Especially, the critical exponents in the scheme II are 
 very close to that of 
 Monte-Calro calculation. \par
 Our aim has been to check whether our numerical method can be reliably 
 applied to the Wilsonian renormalization group equation. 
 We have demonstrated that it works very well in the $O(N)$ linear sigma 
 model. 
 It has been shown also that the method can be applied to finite 
 temperature
 system. This is the first step of our approach to
 more complicated systems including fermions and gauge fields. 
 The Wilsonian RG method provides us with 
 a powerful tool to study non-perturbative features of low energy
 effective theories. 
 As a next step, for instance, 
 this approach may be able to judge whether certain effective model is
 consistent with QCD by comparing
 the solution of the Wilsonian RG equation for the effective model and 
 the result of the lattice QCD. 
%
%
\section*{Acknowledgment}
%
%
 The authors would like to thank Y. Nemoto and M. Tomoyose 
 for fruitful discussions. This work is supported in 
 part by the Grant-in-Aid 
 for Scientific Research (C)(2)08640356 and (C)(2)11640261 of 
 the Ministry of Education, Science,
 Sports and Culture of Japan.
%
%

\end{document}